\pdfoutput=1
\documentclass[final,3p]{elsarticle}

\usepackage{amssymb}

\journal{Computer Physics Communications}

\begin{document}

\begin{frontmatter}



\title{Exact recording of Metropolis-Hastings-class Monte Carlo simulations using one bit per sample}


\author[cmbp,mstp]{Albert H. Mao}
\author[cmbp,bme]{Rohit V. Pappu\corref{author}}

\cortext[author] {Corresponding author.\\\textit{E-mail address:} pappu@wustl.edu}
\address[cmbp]{Computational and Molecular Biophysics Program, Division of Biology and Biomedical Sciences}
\address[mstp]{Medical Scientist Training Program}
\address[bme]{Department of Biomedical Engineering, One Brookings Drive, Campus Box 1097, Washington University in St. Louis, St. Louis, MO 63130-4899}

\begin{abstract}
The Metropolis-Hastings (MH) algorithm is the prototype for a class of Markov chain Monte Carlo methods that propose transitions between states and then accept or reject the proposal.
These methods generate a correlated sequence of random samples that convey information about the desired probability distribution.
Deciding how this information gets recorded is an important step in the practical design of MH-class algorithm implementations.
Many implementations discard most of this information in order to reduce demands on storage capacity and disk writing throughput.
Here, we describe how recording a bit string containing 1's for acceptance and 0's for rejection allows the full sample sequence to be recorded with no information loss, facilitating decoupling of simulation design from the constraints of data analysis.
The recording uses only one bit per sample, which is an upper bound on the rate at which information about the desired distribution is acquired.
We also demonstrate the method and quantify its benefits on a nontrivial colloidal system of charged particles in the canonical ensemble.
The method imposes no restrictions on the system or simulation design and is compatible with descendants of the MH algorithm.
\end{abstract}

\begin{keyword}
Markov chain Monte Carlo \sep Metropolis-Hastings \sep information theory \sep data representation

\PACS 05.10.Ln \sep 07.05.Kf \sep 02.70.Tt

\end{keyword}

\end{frontmatter}


\section{Introduction}

\begin{sloppypar}
Markov chain Monte Carlo (MCMC) methods enable importance sampling from complicated, concentrated probability distributions.
The Metropolis-Hastings (MH) algorithm~\cite{Metropolis1953Equation, Hastings1970Monte} is the prototype for a class of MCMC algorithms where transition proposals are accepted or rejected to generate each sample.
Its applicability to distributions where probability ratios, but not absolute probabilities, are easily computed motivates its broad usage.
Since the complexity and scale of these applications routinely push against computational resource limits, practical techniques for minimizing storage requirements and execution time are important aspects of implementations.
Ideally, the full sequence of samples is recorded in a compact format that facilitates analysis and interpretation.
\end{sloppypar}

Current implementations of Metropolis-Hastings-class algorithms typically reduce the information content of the full sample sequence in two ways before it is recorded.
In the first, a small fraction of samples are recorded in complete detail, but the rest are discarded.
This enables arbitrary post-analysis because all degrees of freedom are available for the retained samples.
However, when sampling systems with many degrees of freedom, retaining even a small fraction of samples requires considerable storage capacity.
Even if capacity is abundant, disk throughput would limit the fraction of samples that could be recorded without having disk write operations dominate the execution time.
In the second pattern, averages, moments, histograms, or other quantities are accumulated during simulation and recorded upon completion.
This is parsimonious with respect to demands on storage bandwidth and capacity, but decisions about which quantities to accumulate, their frequencies of accumulation, and the number of samples to discard due to starting configuration bias must be made in advance.
Altering these choices requires repeating the entire simulation.

These tradeoffs and losses of information complicate the usage of MH-class algorithms in real applications.
A typical experience from our lab provides an illustrative example in the area of biomolecular simulation.
A study of conformational and dimerization equilibria~\cite{Williamson2010Modulation} involved a set of $\sim$$10$ proteins, each consisting of $\sim$$10^3$ interacting atoms and modelled at $\sim$$10$ different temperatures either individually or in pairs.
At least three independent replicate simulations were performed for each condition, with each replicate generating $\sim$$10^8$ samples using MH.
While some quantities of interest had their expectation values accumulated during the initial simulations, only 1 in $\sim$$10^4$ samples had their full set of atomic coordinates recorded.
Despite using Gromacs XTC compression with limited precision, these sparse recordings consumed a total of $\sim$$10^{11}$ bytes of storage.
Subsequent analyses could only be performed by reanalyzing these recorded samples or repeating the entire set of simulations while accumulating expectations for the new quantities of interest.
Given that the simulations consumed $\sim$$10^5$ total hours of cpu time, both options were suboptimal in terms of efficient usage of computational resources.

These problems can be completely avoided.
Here, we describe a simple method for recording the full sample sequence from a MH-class simulation that stores one bit per sample.
The method operates independently from any details of the system and the transition proposals, and is therefore generally applicable.

\section{Description of the method}

Recording a bit string of 1's for acceptance and 0's for rejection, with one bit per transition, suffices to preserve the complete information content of all samples generated during one run of a MH-class algorithm.
This method follows naturally from an information theoretic perspective: since all influence from the underlying distribution is reduced to a binary accept-or-reject decision, MH-class algorithms can be viewed as communication channels with capacity of one bit per sample~\cite[chapter 30.5]{MacKay2002Information}.
An existing implementation of a MH-class algorithm can be easily modified to perform this recording operation during each iteration.
Buffered output is necessary because file systems do not allow writing of individual bits, and also helps to reduce the frequency of disk writes.

Regenerating the full sequence of samples for subsequent reanalysis requires the recorded bit string, the starting state, and any pseudorandom number generator seed(s) from the original simulation.
The original simulation code should be reused to guarantee that the original sequence of transition proposals is recapitulated, but modified to use the recorded bit string for deciding whether to accept or reject proposals.
Since the acceptance criterion no longer needs to be evaluated, all calculations involved in computing ratios of sample weights or proposal probabilities can be skipped during reanalysis.
However, as with the original simulation run, all degrees of freedom for every sample are available for computing and accumulating distributions of any quantity of interest.

Care must be taken to avoid corrupting the sequence of samples during reanalysis.
The stream of pseudorandom numbers generated during reanalysis must be identical to the original simulation's stream.
In particular, if a single pseudorandom number stream is used for generating transition proposals as well as evaluating the acceptance criterion, the pseudorandom number that would have been used for the acceptance criterion must be generated and discarded during each iteration.
If analysis routines themselves make use of pseudorandom numbers, they must generate their own independent streams.
Insidious platform dependencies are another source of potential corruption when recordings generated on one computer are reanalyzed on another.
For example, the implementation of floating point arithmetic differs between processor architectures and compilers.
This source of error should be eliminated by adhering to best practices in the coding of floating point operations~\cite{Goldberg1991What} or writing unit tests that assert platform-specific assumptions are valid during both original simulation and reanalysis runs.

\section{Demonstration of the method}

To create a nontrivial demonstration of this recording method, we implemented a MH simulation of a colloidal suspension of charged spherical particles.
This three-dimensional off-lattice system consists of $10^3$ particles confined within a spherical droplet.
Each particle has two charge sites that freely diffuse on the particle's surface.
The potential energy is the sum of a pairwise Lennard-Jones potential between particle centers and a pairwise Debye-H\"{u}ckel screened electrostatic potential between charge sites.
Intra-particle and inter-particle electrostatic interactions are screened using different dielectric constants.
Transition proposals consist of local or full randomization of one particle's center coordinates or charge site positions.
An executable Java Archive file containing source code and compiled class files for this demonstration is available as supplementary material.
To run it on any computer where Java Runtime Environment version 6 is installed, execute \texttt{java~-jar~RecordingDemonstration.jar} at a command line.

A comparison with typical alternatives highlights the efficiency of the proposed recording method.
A straightforward format would be a recording of all seven degrees of freedom $(x, y, z, \theta_1, \phi_1, \theta_2, \phi_2)$ for all particles.
While this is obviously inefficient, it is simple and easy to parse, facilitating interoperability with other software.
Storing the updated values, if any, for only the changed degrees of freedom during a transition would be significantly more efficient.
However, this method would tie the recording format to the choice of transition proposals; modifying the simulation to use more sophisticated transitions that simultaneously perturb multiple particles would require redesigning the format.
An even more efficient method would record the potential energy (or more generally, the weight) of each sample.
As with the proposed method, this would require preserving the simulation code, but would enable full reconstruction of the original samples without calculating any weights.
Table~\ref{table:storagetable} compares the storage efficiency of these recording schemes to the proposed one.
\texttt{RecordingDemonstration.jar} can derive any of the other recordings starting from the bit string recording, proving that the bit string (along with simulation code and starting state) retains all information about the sample sequence.

\begin{table}[htbp]
\centering
\begin{tabular}{l p{0.5\linewidth} r}

\hline
& Method & Bytes per sample \\
\hline

1 & All degrees of freedom for all particles & 56000\phantom{.000} \\
2 & All degrees of freedom for changed particle only & 32\phantom{.000} \\
3 & Energy output & 8\phantom{.000} \\
4 & Bit string & 0.125 \\

\hline
\end{tabular}
\caption{
Comparison of efficiencies for different recording methods.
Calculated sizes assume that real numbers are represented and recorded using an eight-byte format.
Method 2 uses one four-byte integer per sample to encode which particle, if any, changed, and assumes an acceptance rate of exactly 50\%.
Note that preserving the simulation code is necessary for methods 3 and 4.
This would consume an additional amount of storage that is independent of the number of samples, and is not reflected in the table.
}
\label{table:storagetable}
\end{table}

\begin{sloppypar}
Since the bit string recording obviates all sample weight and proposal probability ratio calculations, iterating over the sample sequence takes significantly less time than generating it.
To enable benchmarking, \texttt{RecordingDemonstration.jar} provides two analyses for the demonstration system.
The first is a histogram of the central angle between intra-particle charge sites and the second is a pairwise distance histogram between particle centers.
The histograms are accumulated once every 100 and 1000 steps, respectively.
Table~\ref{table:timingtable} compares the execution time for performing both analyses during the original simulation versus using the bit string recording.
Note that energy evaluations are efficiently implemented such that only changing terms are computed; each iteration after the first of the original simulation performs a number of computations that is linear, rather than quadratic, in the number of particles.
However, once the bit string is recorded, energy evaluations are skipped; each iteration only needs to update the degrees of freedom for a single particle and therefore executes in constant time.
\end{sloppypar}

\begin{table}[htbp]
\centering
\begin{tabular}{p{0.5\linewidth} r}

\hline
Operation & Total running time \\
 & (seconds) \\
\hline

Record bit string and analyze & 2809 \\
Record bit string only & 2773 \\
Reanalyze using bit string & 85 \\

\hline
\end{tabular}
\caption{
Comparison of total run times for generating $10^6$ samples of the demonstration system.
Benchmarks were performed on a 2.6 GHz Intel Core 2 Duo system with 4 GB of 667 MHz DDR2 SDRAM and 6 MB L2 cache running version 1.6.0\_22 of the Java Runtime Environment.
}
\label{table:timingtable}
\end{table}

\section{Discussion}

This method enables efficient reanalysis as long as the time spent generating and executing transition proposals is small compared to the time spent computing ratios of sample weights and proposal probabilities.
Fortunately, this condition is naturally satisfied because sample weights are generally determined by an energy function or other interaction between degrees of freedom that is more expensive to update than the degrees of freedom themselves.
By recording the results of the most computationally expensive component of MH-class algorithm implementations, the method proposed here approaches the minimum achievable limits on both storage consumption and execution time.

A closely related alternative method would be to record the full sequence of sample weights in addition to the accept / reject decision.
In some applications, the sample weights themselves are important subjects of analysis, and rederiving them from the samples during reanalysis would be computationally expensive.
For instance, in a simulation of a physical system in the canonical ensemble, one may wish to calculate the heat capacity and other properties of the energy distribution.
Assuming that weights are represented as IEEE 754 double precision floating point numbers, as in Table~\ref{table:storagetable}, each sample would require an additional 64 bits of storage.
While much less compact than the bit string recording alone, it would still be far more efficient than most alternatives.

Note that the size of the recording might be further reduced by compressing it using a lossless algorithm.
One bit per sample is an \emph{upper} bound on the entropy rate~\cite[chapter 30.5]{MacKay2002Information}; if the acceptance rate of the simulation is not exactly 1/2, the recorded bit string will contain biases that a compression algorithm can exploit.

The compactness of this recording method makes it susceptible to corruption: during reanalysis, a single incorrectly recorded bit contaminates all subsequent samples.
An effective solution would be to encode the raw bit string using error correcting codes (ECC) that provide robustness at the cost of increased storage consumption~\cite[chapter 11]{MacKay2002Information}.
In fact, employment of Reed-Solomon~\cite{Reed1960Polynomial} and low-density parity-check (LDPC)~\cite{Gallager1962Lowdensity} codes at the hardware level is already widespread among designers of random access memory, persistent storage devices, and networking interfaces.
Therefore, ECC at the software level would constitute a redundant layer of protection against error.
Non-redundant detection of error can be achieved by recording all degrees of freedom for a small number of checkpoint samples, and verifying equality for the corresponding samples generated during reanalysis.
Alternatively, the common practice of comparing cryptographic hash function outputs can be applied to verifying the integrity of recorded bit strings.

This method is compatible with descendants of the MH algorithm that retain its general accept-or-reject architecture.
Examples include expanded ensemble techniques~\cite{Lyubartsev1992New}, replica exchange Monte Carlo~\cite{Mitsutake2001Generalizedensemble}, multiple-try Metropolis~\cite{Liu2000MultipleTry}, simulated annealing~\cite{Kirkpatrick1983Optimization}, and simulated tempering~\cite{Marinari1992Simulated}.
For some algorithms, such as Wang-Landau sampling~\cite{Wang2001Efficient}, efficient post-analysis would require recording the system energy at every iteration as described above.

\section{Conclusion}

The method described here is a simple and effective approach to data storage and representation in the design of MH-class algorithm implementations.
It facilitates decoupling of simulation design from the constraints of data analysis.
The information theoretic perspective through which this method was conceived deserves broader appreciation in the development of Monte Carlo algorithms.

\section{Acknowledgements}

This work was supported by National Science Foundation MCB 0718924 and National Institutes of Health – National Institute of General Medical Sciences 5T32GM008802.

%





\bibliographystyle{model1-num-names}
\bibliography{bibliography}

\begin{thebibliography}{13}
\expandafter\ifx\csname natexlab\endcsname\relax\def\natexlab#1{#1}\fi
\providecommand{\bibinfo}[2]{#2}
\ifx\xfnm\relax \def\xfnm[#1]{\unskip,\space#1}\fi
\bibitem[{Metropolis et~al.(1953)Metropolis, Rosenbluth, Rosenbluth, Teller,
  and Teller}]{Metropolis1953Equation}
\bibinfo{author}{N.~Metropolis}, \bibinfo{author}{A.~W. Rosenbluth},
  \bibinfo{author}{M.~N. Rosenbluth}, \bibinfo{author}{A.~H. Teller},
  \bibinfo{author}{E.~Teller},
\newblock \bibinfo{title}{{Equation of State Calculations by Fast Computing
  Machines}},
\newblock \bibinfo{journal}{The Journal of Chemical Physics}
  \bibinfo{volume}{21} (\bibinfo{year}{1953}) \bibinfo{pages}{1087--1092}.
\bibitem[{Hastings(1970)}]{Hastings1970Monte}
\bibinfo{author}{W.~K. Hastings},
\newblock \bibinfo{title}{{Monte Carlo sampling methods using Markov chains and
  their applications}},
\newblock \bibinfo{journal}{Biometrika} \bibinfo{volume}{57}
  (\bibinfo{year}{1970}) \bibinfo{pages}{97--109}.
\bibitem[{Williamson et~al.(2010)Williamson, Vitalis, Crick, and
  Pappu}]{Williamson2010Modulation}
\bibinfo{author}{T.~E. Williamson}, \bibinfo{author}{A.~Vitalis},
  \bibinfo{author}{S.~L. Crick}, \bibinfo{author}{R.~V. Pappu},
\newblock \bibinfo{title}{{Modulation of Polyglutamine Conformations and Dimer
  Formation by the N-Terminus of Huntingtin}},
\newblock \bibinfo{journal}{Journal of Molecular Biology} \bibinfo{volume}{396}
  (\bibinfo{year}{2010}) \bibinfo{pages}{1295--1309}.
\bibitem[{MacKay(2002)}]{MacKay2002Information}
\bibinfo{author}{D.~J.~C. MacKay}, \bibinfo{title}{{Information Theory,
  Inference \& Learning Algorithms}}, \bibinfo{publisher}{Cambridge University
  Press}, \bibinfo{edition}{1st} edition, \bibinfo{year}{2002}.
\bibitem[{Goldberg(1991)}]{Goldberg1991What}
\bibinfo{author}{D.~Goldberg},
\newblock \bibinfo{title}{{What every computer scientist should know about
  floating-point arithmetic}},
\newblock \bibinfo{journal}{ACM Computing Surveys} \bibinfo{volume}{23}
  (\bibinfo{year}{1991}) \bibinfo{pages}{5--48}.
\bibitem[{Reed and Solomon(1960)}]{Reed1960Polynomial}
\bibinfo{author}{I.~S. Reed}, \bibinfo{author}{G.~Solomon},
\newblock \bibinfo{title}{{Polynomial Codes Over Certain Finite Fields}},
\newblock \bibinfo{journal}{Journal of the Society for Industrial and Applied
  Mathematics} \bibinfo{volume}{8} (\bibinfo{year}{1960})
  \bibinfo{pages}{300--304}.
\bibitem[{Gallager(1962)}]{Gallager1962Lowdensity}
\bibinfo{author}{R.~Gallager},
\newblock \bibinfo{title}{{Low-density parity-check codes}},
\newblock \bibinfo{journal}{IEEE Transactions on Information Theory}
  \bibinfo{volume}{8} (\bibinfo{year}{1962}) \bibinfo{pages}{21--28}.
\bibitem[{Lyubartsev et~al.(1992)Lyubartsev, Martsinovski, Shevkunov, and
  Velyaminov}]{Lyubartsev1992New}
\bibinfo{author}{A.~P. Lyubartsev}, \bibinfo{author}{A.~A. Martsinovski},
  \bibinfo{author}{S.~V. Shevkunov}, \bibinfo{author}{P.~N.~V. Velyaminov},
\newblock \bibinfo{title}{{New approach to Monte Carlo calculation of the free
  energy: Method of expanded ensembles}},
\newblock \bibinfo{journal}{The Journal of Chemical Physics}
  \bibinfo{volume}{96} (\bibinfo{year}{1992}) \bibinfo{pages}{1776--1783}.
\bibitem[{Mitsutake et~al.(2001)Mitsutake, Sugita, and
  Okamoto}]{Mitsutake2001Generalizedensemble}
\bibinfo{author}{A.~Mitsutake}, \bibinfo{author}{Y.~Sugita},
  \bibinfo{author}{Y.~Okamoto},
\newblock \bibinfo{title}{{Generalized-ensemble algorithms for molecular
  simulations of biopolymers}},
\newblock \bibinfo{journal}{Biopolymers} \bibinfo{volume}{60}
  (\bibinfo{year}{2001}) \bibinfo{pages}{96--123}.
\bibitem[{Liu et~al.(2000)Liu, Liang, and Wong}]{Liu2000MultipleTry}
\bibinfo{author}{J.~S. Liu}, \bibinfo{author}{F.~Liang}, \bibinfo{author}{W.~H.
  Wong},
\newblock \bibinfo{title}{{The Multiple-Try Method and Local Optimization in
  Metropolis Sampling}},
\newblock \bibinfo{journal}{Journal of the American Statistical Association}
  \bibinfo{volume}{95} (\bibinfo{year}{2000}) \bibinfo{pages}{121--134}.
\bibitem[{Kirkpatrick et~al.(1983)Kirkpatrick, Gelatt, and
  Vecchi}]{Kirkpatrick1983Optimization}
\bibinfo{author}{S.~Kirkpatrick}, \bibinfo{author}{C.~D. Gelatt},
  \bibinfo{author}{M.~P. Vecchi},
\newblock \bibinfo{title}{{Optimization by Simulated Annealing}},
\newblock \bibinfo{journal}{Science} \bibinfo{volume}{220}
  (\bibinfo{year}{1983}) \bibinfo{pages}{671--680}.
\bibitem[{Marinari and Parisi(1992)}]{Marinari1992Simulated}
\bibinfo{author}{E.~Marinari}, \bibinfo{author}{G.~Parisi},
\newblock \bibinfo{title}{{Simulated Tempering: A New Monte Carlo Scheme}},
\newblock \bibinfo{journal}{Europhysics Letters} \bibinfo{volume}{19}
  (\bibinfo{year}{1992}) \bibinfo{pages}{451--458}.
\bibitem[{Wang and Landau(2001)}]{Wang2001Efficient}
\bibinfo{author}{F.~Wang}, \bibinfo{author}{D.~P. Landau},
\newblock \bibinfo{title}{{Efficient, Multiple-Range Random Walk Algorithm to
  Calculate the Density of States}},
\newblock \bibinfo{journal}{Physical Review Letters} \bibinfo{volume}{86}
  (\bibinfo{year}{2001}) \bibinfo{pages}{2050--2053}.

\end{thebibliography}







\end{document}